\documentclass[twocolumn,aps,10pt,showpacs,showkeys,]{revtex4}
\usepackage[dvips]{graphicx}

\begin{document}
\title{Electronic structure and magnetism
of LaVO$_3$ and LaMnO$_3$$^\spadesuit$}
\author{R. J. Radwanski}
\affiliation{Center of Solid State Physics, S$^{nt}$Filip 5, 31-150 Krakow, Poland,\\
Institute of Physics, Pedagogical University, 30-084 Krakow,
Poland}
\author{Z. Ropka}
\affiliation{Center of Solid State Physics, S$^{nt}$Filip 5,
31-150 Krakow, Poland}
\homepage{http://www.css-physics.edu.pl}
\email{sfradwan@cyf-kr.edu.pl}

\begin{abstract}
In this contribution we derive and discuss energy levels of the
  strongly-correlated d$^2$ configuration of the V$^{3+}$ ion (LaVO$_3$) and of
  d$^4$ configuration of the Mn$^{3+}$ ion (LaMnO$_3$) in the octahedral surroundings in the
  presence of the spin-orbit coupling and the resulting magnetic
  properties. We take into account very strong correlations among
the d electrons and work with strongly-correlated atomic-like
electronic systems, ground term of which is, also in a solid,
described by two Hund's rules quantum numbers. In a solid we take
into account the influence of crystal-field interactions,
predominantly of the cubic (octahedral) symmetry. We describe both
paramagnetic state and the magnetically-ordered state getting a
value of 1.3 $\mu_B$ for the V$^{3+}$-ion magnetic moment in the
ordered state at 0 K for LaVO$_3$ ($^3$T$_{1g}$) and of 3.7
$\mu$$_B$ for LaMnO$_3$ ($^5$E$_g$). A remarkably consistent
description of both zero-temperature properties and thermodynamic
properties indicates on the high physical adequacy of the applied
atomic approach. We point out the necessity to unquench the
orbital moment in 3d-ion compounds. We claim that the
intra-atomic spin-orbit coupling and strong electron correlations
are indispensable for the physically adequate description of
electronic and magnetic properties of LaVO$_{3}$ and LaMnO$_{3}$.

\end{abstract}
\maketitle

\section{Introduction and the aim}
\vspace {-0.4 cm}
 LaVO$_{3}$ and LaMnO$_{3}$ attract a great
scientific interest by more than 50 years [1,2]. Despite of it
there is still strong discussion about the description of its
properties and its electronic structure. A controversy starts
already with the description of the electronic ground state. For
V$_2$O$_3$, a canonical compound of LaVO$_{3}$ there is within a
localized paradigm a long-standing controversy between a {\it S}=1
model without an orbital degeneracy [3] and the historically first
$S$=1/2 orbitally degenerate model of Castellani et al.[1]. A
spin-1 model with three degenerate orbitals was worked out by Di
Matteo [4] whereas Refs [5,6] develop a model where the
fluctuations of t$_{2g}$ orbitals and frustrations play prominent
role in magnetism of LaVO$_3$.

In all of these considerations it is agreed that LaVO$_{3}$ is an
insulating antiferromagnet with the Neel temperature of 135-140 K.
LaMnO$_{3}$ is also insulating antiferromagnet with T$_N$ of 140
K. The basis for all theories is the description of the V$^{3+}$
ion (2 d electrons) and the Mn$^{3+}$ ion (4 d electrons) and
their electronic structure in the perovskite crystallographic
structure. In the perovskite-based structure of LaVO$_3$ and
LaMnO$_3$ the V and Mn ions are placed in a distorted oxygen
octahedron.

The aim of this paper is to present a consistent description of
the low-energy electronic structure and of the magnetism of
LaVO$_{3}$ and LaMnO$_{3}$ as originating from the atomic-like
electronic structure of the strongly-correlated 3$d^{2}$ and
3$d^{4}$ electronic systems occurring in the V$^{3+}$ and
Mn$^{3+}$ ions. In our description local atomic-scale effects, the
orbital magnetism and the intra-atomic spin-orbit coupling play
the fundamentally important role in determination of properties
of the whole compound.

\section{$\textrm{d}$ states of the V$^{3+}$ ion in atomic physics}
\vspace {-0.4 cm} According to our approach to 3d compounds, that
seems to be very natural, but is far from being accepted within
the magnetic community, two $d$ electrons of the V$^{3+}$ ion in
the incomplete 3$d$ shell in LaVO$_{3}$ form the strongly
correlated intra-atomic 3$d^{2}$ electron system. These strong
correlations among the 3$d$ electrons we account for by two
Hund's rules, that yield the $^{3}F$ ground term, Fig. 1a. In the
oxygen octahedron surroundings, realized in the perovskite
structure of LaVO$_{3}$, the $^{3}F$ term splits into two orbital
triplets $^{3}$T$_{1g}$, $^{3}$T$_{2g}$ and an orbital singlet
$^{3}A_{2g}$. In the octahedron with negative oxygen ions the
ground subterm is orbital triplet $^{3}$T$_{1g}$ as is shown in
Fig. 1b. All orbital states are triply spin degenerated. This
degeneration is removed by the intra-atomic spin-orbit coupling
that is always present in the ion/atom. The effect of the
spin-orbit coupling is shown in Fig. 1c. For low-temperature
properties the lowest 5 states of nine states of the
$^{3}$T$_{1g}$ subterm are important. Off-octahedral distortions
causes further splitting of the shown states - important is that
there is no more low-energy states as only shown. The present
calculations have been performed with a realistic octahedral CEF
parameter B$_4$= -40 K (=10Dq =600B$_4$= 2.06 eV) and the
spin-orbit coupling as is observed in the free ion, i.e.
$\lambda_{s-o}$= +150 K. The spin-orbit coupling effect amounts
only to 3$\%$ the CEF effect but it has enormous influence on the
eigen-functions and low-temperature properties.
\begin{figure}[t]
\begin{center}
\includegraphics[width = 8.5 cm]{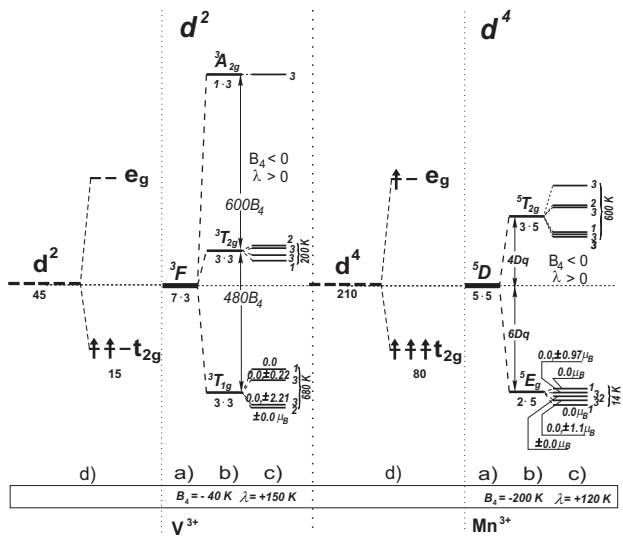}
\end{center} \vspace {-0.3 cm}
\caption{Lowest part of the electronic structure of the d$^2$
(V$^{3+}$ ion) in LaVO$_3$ and of the d$^4$ configuration
(Mn$^{3+}$ ion) in LaMnO$_3$ in the strongly-correlated limit. d)
schematic electronic structures customarily recalled in the
current literature - these single-electron crystal-field states
are, according to us, completely oversimplified. a) Hund's rules
ground terms, b) - effect of the octahedral CEF interactions, c)
- the many-electron CEF approach with strong intra-atomic
correlations propagating by us. c) many-electron CEF states in
the octahedral CEF in the presence of the intra-atomic spin-orbit
coupling.}
\end{figure}

The derived many-electron electronic structure of the d$^2$
configuration is completely different from very simplified
electronic structure, shown in Fig. 1d, usually presented in the
current literature with two parallel spin-electrons in the triply
degenerated t$_{2g}$ orbitals. We do not know, why such an
oversimplified scheme is usually shown despite of the fact that
the many-electron CEF approach is known already by almost 70
years, though it is true that it was rather used not to a solid,
but to 3d ions as impurities in Electron Paramagnetic Resonance
(EPR) experiments [7]. For completeness of the description of the
d$^2$ configuration we can add that for two $d$ electrons there
is in total 45 states that in the free V$^{3+}$ ion are arranged
into 5 terms ($^3F$, $^1D$, $^3P$, $^1G$ and $^1S$). The 21-fold
degenerated $^3F$ term is the lowest one according to two Hund's
rules ($S$=1, $L$=3) - it is a reason that its splitting we show
in Fig. 1. The octahedral CEF splits these 45 states in 5 terms
into 11 subterms as has been calculated almost 50 years ago by
Tanabe and Sugano [8]. A $t^2_{2g}$ state, usually recalled in the
literature, is related to the very strong crystal-field limit
notation and is 15-fold degenerated. It contains 4 subterms:
$^3$T$_{1g}$, $^1$T$_{2g}$, $^1$E$_g$, $^1$A$_{1g}$ with the
lowest $^3$T$_{1g}$ subterm. In our approach, Quantum Atomistic
Solid-State Theory, to 3d-ion compounds we assume the CEF to be
stronger than the spin-orbit coupling, but not so strong to
destroy intra-atomic correlations. Thus we are working in an
intermediate crystal-field limit.

\section{Influence of the spin-orbit coupling on low-temperature susceptibility}
\vspace {-0.4 cm}
 In Fig. 2 we present results of our
calculations for the paramagnetic susceptibility for different
physical situations, without the spin-orbit coupling and with and
without octahedral crystal field, which are shown in the
$\chi^{-1}$ $\it {vs}$ T plot. These two curves, without the
spin-orbit coupling, resemble the Curie law but with the
paramagnetic effective moment much larger than expected for the
free $S$=1 spin, of 2.82 $\mu_B$. An introduction of the
spin-orbit interactions causes dramatic change of the $\chi^{-1}$
$\it {vs}$ T plot - the Curie law is broken. If one would like to
analyze this $\chi^{-1}$ $\it {vs}$ T plot within the conventional
paradigm would infer a Curie-Weiss law with an appearance of AF
interactions with $\theta_{CW}$ of about 500 K. However, such the
C-W behaviour appears here as the effect of the intra-atomic
spin-orbit coupling.
\begin{figure}[t]
\begin{center}
\includegraphics[width = 8.9 cm]{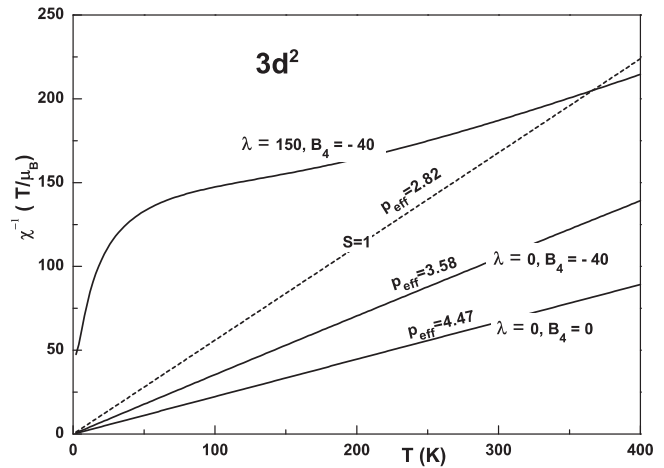}
\end{center} \vspace {-0.5cm}
\caption{Calculated temperature dependence of the paramagnetic
susceptibility for the strongly-correlated d$^2$ configuration
for different physical situations. A dramatic effect of the
spin-orbit coupling on the $\chi^{-1}$ $\it {vs}$ T dependence
should be noticed. }
\end{figure}

\section{Magnetic state of L$\textrm{a}$VO$_3$}
\vspace {-0.4 cm}
 For calculations of the magnetic state apart of
the a single-ion like CEF Hamiltonian we take into account
additionally intersite spin-dependent interactions (n$_{dd}$ -
molecular field coefficient, magnetic moment $m_d$ =
-(L+2S)$\mu_B$) [9]:
\\

H=$H_{CF}$+$H_{d-d}$=
\begin{equation}
  \label{eq:1}
  \sum \sum B_n^mO_n^m+n_{dd}\left(m_d
\left\langle m_d\right\rangle -\frac 12\left\langle
m_d\right\rangle ^2\right)\;.
\end{equation}

In Equation (1) the first term is the crystal-field Hamiltonian:
here we work with B$_4$= -40 K (an ordinary point-charge
octahedron with the V-O distance of 2.0 $\dot{A}$ gives B$_4$= -18
K) and a small tetragonal distortion B$_2$= -60 K. The second
term takes into account intersite spin-dependent interactions
that is necessary to produce the magnetic order below T$_N$. The
self-consistent calculations have been performed similarly to
that presented in Ref. [9] for FeBr$_{2}$. The calculated
temperature dependence of the magnetic moment is shown in Fig. 3.
A value of 1.3 $\mu_ B $ and $T_N$ of 140 K are in perfect
agreement with experimental observations. We have to add,
however, that due to 5 closely lying discrete states the V$^{3+}$
ion is very susceptible to lattice distortions and details of
magnetic interactions.
\begin{figure}[t]
\begin{center}
\includegraphics[width = 8.8 cm]{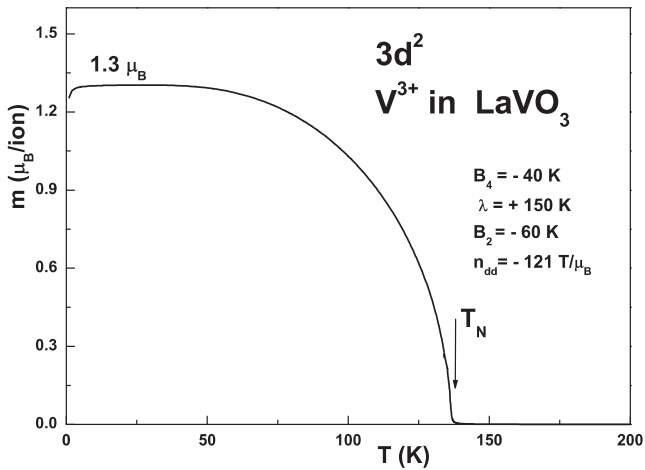}
\end{center}\vspace {-0.5 cm}
\caption{Temperature dependence of the magnetic moment of the
V$^{3+}$ ion in LaVO$_3$. An anomaly at low temperatures is due
to an accidental double degeneracy. }
\end{figure}
\section{$\textrm{d}$ states of the M$\textrm{n}$$^{3+}$ ion in atomic physics}
\vspace {-0.4 cm}
 In our view, the four $d$ electrons of the
Mn$^{3+}$ ion in the incomplete 3$d$ shell in LaMnO$_{3}$ form
the strongly correlated intra-atomic 3$d^{4}$ electron system.
These strong correlations among the 3$d$ electrons we account for
by two Hund's rules, that yield the $^{5}D$ ground term, Fig. 1a,
with $S$=2 and $L$=2. In the oxygen octahedron surroundings,
realized in the perovskite structure of LaMnO$_{3}$, the $^{5}D$
term splits into the orbital doublet $^{5}E_{g}$ as the ground
subterm, Fig. 1b, and the excited orbital triplet $^{5}$T$_{2g}$.

The $E_{g}$ ground subterm comes out from \textit{ab initio}
calculations for the octupolar potential, the A$_{4}$ CEF
coefficient acting on the Mn$^{3+}$ ion from the oxygen negative
charges. Such the atomic-like 3$d^{4}$ system interacts with the
charge and spin surroundings in the solid.

In Ref. 10 we have described LaMnO$_3$ with the octahedral CEF
parameter $B_{4}$ = -13 meV, $B_{2}^{0}$=+10 meV, the spin-orbit
coupling parameter $\lambda _{s-o}$=+33 meV and the
molecular-field coefficient n=-26.3 T/$\mu _{B}$. We get a value
of 3.72 $\mu _{B}$ for the Mn$^{3+}$-ion ordered magnetic moment
that orders magnetically along the $a$ axis within the tetragonal
plane. The magnetic interactions set up at 0 K the molecular
field of 108 T (in case of LaVO$_3$ we get 156 T). The calculated
magnetic moment of 3.72 $\mu _{B}$ is built up from the spin
moment of +3.96 $\mu _{B}$ and from the orbital moment of -0.24
$\mu _{B}$. Although the orbital moment is relatively small its
presence modifies completely the electronic structure similarly
as occurs in case of the V$^{3+}$ ion.

We advocate by years for the importance of the
crystal-field-based approach to transition-metal compounds. Of
course, we do not claim that CEF explains everything (surely CEF
itself cannot explain the formation of a magnetic ordering, for
which in Eq. 1 we add the second term), but surely the
atomic-like 4f/5f/3d cation is the source of the magnetism of a
whole solid. The atomic-scale magnetic moment is determined by
local effects known as crystal field and spin-orbit interactions
as well as very strong electron correlations. These strong
electron correlations are predominantly of the intra-atomic
origin and are taking into account, in the first approximation,
via on-site Hund's rules. These strong correlations lead to
many-electron version of the CEF approach in contrary to the
single-electron version of the CEF presently popular.

Finally, we use a name "$d$ electrons", as often is used in
literature. Of course, $d$ electrons are not special electrons
but it means electrons in $d$ states. $d$ states have well
defined characteristics, like the orbital quantum number $l$. By
identifying states in a solid we can identified $d$ electrons, in
particular their number.

\section{Conclusions}
\vspace {-0.4 cm}
 We have calculated electronic structures of
d$^2$ and d$^4$ systems occurring in V$^{3+}$ and Mn$^{3+}$ ions
in LaVO$_3$ and LaMnO$_3$ taking into account strong electron
correlations and the spin-orbit coupling. On basis of our studies
we reject a possibility for an orbital liquid both in LaVO$_3$
and LaMnO$_3$. Both ions are strongly susceptible to lattice
distortions. We reproduce a value of 1.3 $\mu _{B}$ for the
V$^{3+}$-ion magnetic moment as well as the Neel temperature. We
advocate for years for an atomistic crystal-field approach to
3d-ion compounds with a discrete electronic structure and the
insulating ground state as an alternative to the band picture
[11] knowing its enormous shortage in description of 3d oxides
(metallic instead ionic ground state, for instance). Our approach
accounts both for the insulating ground state, magnetism as well
as thermodynamical properties.\\
$^\spadesuit$ Dedicated to the Pope John Paul II, a man of
freedom and truth in life and in Science.

\end{document}